\documentclass[floatfix,10pt,superscriptaddress,twocolumn,showpacs,showkeys,aps,prb,final]{revtex4-1}

\usepackage{graphicx}
\usepackage{dcolumn}
\usepackage{subfigure}
\usepackage{bm}
\usepackage{ifpdf,braket}
\usepackage{epstopdf}

\ifpdf
\usepackage[pdftex,
        colorlinks=true,
        pdftitle={Variational principle applied to two-band superconductors},
        pdfauthor={Antonio R. de C. Romaguera},
        pdfsubject={Vortex theory},
        pdfkeywords={Departamento de Fisica, Universidade Federal Rural de Pernambuco, 52171-900 Recife, Pernambuco, Brazil},
        baseurl={http://www.cmt.ua.ac.be},
        pdftoolbar=true,
        pdfmenubar=false,
        pdfpagemode=UseOutlines,
        pdfstartview=FitH,
        bookmarksopen=false
        ]{hyperref}
\fi

\begin{document}

\title{Topologically stable gapped state in a layered superconductor}
\author{Marco Cariglia}%
\affiliation{Departamento de F\'{\i}sica , Universidade Federal de
Ouro Preto, 35400-000 Ouro Preto Minas Gerais, Brazil}
\author{Alfredo A. Vargas-Paredes}%
\affiliation{Departamento de F\'{\i}sica dos S\'{o}lidos,
Universidade Federal do Rio de Janeiro, 21941-972 Rio de Janeiro,
Brazil}
\author{Mauro M. Doria}%
\affiliation{Departamento de F\'{\i}sica dos S\'{o}lidos, Universidade Federal do Rio de Janeiro, 21941-972 Rio de Janeiro, Brazil}%
\email{mmd@if.ufrj.br}

\date{\today}

\begin{abstract}
We show that a layered superconductor, described by a two-component
order parameter, has a gapped state above the ground state,
topologically protected from decay, containing flow and counter flow
in the absence of an applied magnetic field. This state is made of
skyrmions, breaks time reversal symmetry and produces a weak local
magnetic field below the present threshold of detection by $\mu$SR
and NMR/NQR. We estimate the density of carriers that condense into
the pseudogap.
\end{abstract}


\maketitle

\section{Introduction}

An important concept in condensed matter
physics is that of an order parameter (OP), introduced by Lev Landau
in the last century to describe the transition to the
superconducting state. Interestingly, one of Landau's first
proposals of an order parameter was the supercurrent, also suggested
to exist in the microscopic superconducting ground state of Felix
Bloch~\cite{schmalian10}. These ideas were soon dismissed since a
spontaneous circulating supercurrent increases the kinetic energy.
In this letter we find an excited but stable state with supercurrents, containing flow and counter flow in the layers, even without the presence of an external magnetic field. This is a skyrmion state found to exist in a layered superconductor described by a two-component OP. The decay of the skyrmion state into other configurations of lower free energy is prevented by its topological stability, which gives rise to an energy gap since it lies above  the homogeneous state. The skyrmion state breaks the time reversal symmetry and we find that the very weak magnetic field, produced by the spontaneously circulating supercurrents, determines the value of the gap above the homogeneous state. The gap above the ground state, the topological stability and the unusual magnetic order that breaks the time reversal symmetry found here leads us to suggest that the pseudogap of the layered superconductors is indeed a skyrmion state.

We derive the OP and the local magnetic field associated to the skyrmion state from the {\it first order equations} (FOE) instead of the second order variational equations. Interestingly A. A. Abrikosov
\cite{abrikosov57}, used the FOE to discover the vortex lattice  instead of the second order Ginzburg-Landau (GL) variational equations. The GL free energy only sorts among the possible vortex lattice the one with minimal energy. Later the FOE were rediscovered by E. Bogomolny~\cite{bogomolny76} in the context of string theory and shown to solve exactly  the GL second order equations for a particular value of the coupling constant. The Seiberg-Witten equations, which also describe topological excitations, namely four dimensional monopoles~\cite{seiberg94}, are FOE that belong to the same family of the Abrikosov-Bogomolny (AB) equations. In this family, the FOE determine the OP and the vector potential associated to topological excitations. In
this letter we obtain the topological solutions (skyrmions) associated to another set of equations that belong to this family that lives in the three-dimensional Euclidean space~\cite{doria10, alfredo13}. The existence of topological solutions for the two-component GL theory was pointed out long ago~\cite{babaev02}, through a mapping into a nonlinear O(3) sigma model. Recently skyrmion solutions were obtained from the second order equations of the two-component GL theory, but only in the two-dimensional limit, due to an uniaxial symmetry along the direction of an applied field~\cite{garaud12}. Here we obtain skyrmions as truly three-dimensional topological solutions of the FOE of the two-component GL theory, applied to a stack of layers and without the presence of an applied  magnetic field. We find that the skyrmions form an excited gapped state regardless of the shape of the condensate energy.

\section{The skyrmion state}
In his original work T. Skyrme made
protons and neutrons stable by association to a topologically
non-trivial solution of the sigma model~\cite{skyrme62,
kalbermann91}, the skyrmion, that represents a configuration with
particle-like properties. Skyrmions are found in many condensed
matter systems with an inner structure constructed over a physical OP.
Skyrmions were reported in the quantized Hall
effect~\cite{schmeller95}, Bose-Einstein
condensates~\cite{volovik12} and
superfluid$^3$He-A~\cite{walmsley12}. In the antiferromagnet
La$_2$Cu$_{1-x}$Li$_x$O$_4$~\cite{marcelo11} the CuO$_2$ layers are
modified by Li atoms, which are dopants that frustrate the original
Neel state, leading to a magnetic state made of skyrmions. Recently
skyrmions were found to form a crystalline order~\cite{yu10} in the
helimagnet MnSi~\cite{pfleiderer04} and also in the doped
semiconductor Fe$_{1-x}$Co$_x$Si~\cite{munzer10}.

The present skyrmion state generates a  magnetic field $\vec h$ between the layers due to the superficial supercurrent $\vec J_s$ circulating in the layers. In this letter we assume a tetragonal symmetric skyrmion state, described by a unit cell of size $L$.
We find here that the skyrmion state exists around the minimum of the magnetic energy density, $\vec h^2/8\pi$, stored between layers, which takes place for a special value of the ratio between the unit cell size and the distance between consecutive layers, $d$. Surprisingly our theoretically found ratio $L/d$ coincides with the experimentally obtained ratio $L'/d$~\cite{edinardo13}, where $L'$ is the distance between dopants within a copper-oxygen layer. According to R.P. Roeser et al.~\cite{roeser08} the tetragonal structure set by $L'$ at maximum doping acts as a resonator that we claim here to be that of the skyrmion state, namely, $L'=L$.
Indeed since the discovery of stripe order in the
cuprates, charge, magnetism and superconductivity are believed to
coexist in geometrical periodic arrangements \cite{tranquada08}.

\begin{figure}[htb]
\includegraphics[width=\linewidth]{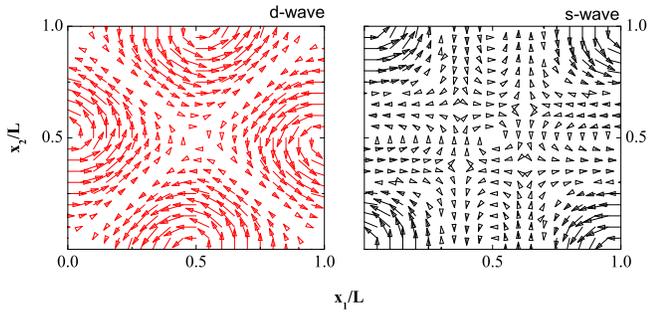}
\caption{\label{currents}(color online) The superficial current
$\vec J_s$ is shown for d and s waves for a square unit cell. The
two d-wave skyrmions are centered in the middle of the sides and the
single s-wave skyrmion is at the corner.}
\end{figure}

The skyrmion lattice associated to d and s wave symmetrical OP are shown in fig.~(\ref{currents}). There are two cores of circulating currents for d wave (red), with centers in the middle of the unit cell sides, while for s wave (black) there is only one core centered at the corners of the unit cell. In both cases, there is no net circulation in the unit cell, and yet there is a single sense of intense circulation set in the cores. As seen from above a layer, e.g. the plane of fig.~(\ref{currents}), the $\vec h$ stream lines
sink into these cores to re-emerge from below, and pierce the unit
cell a second time, forming closed loops. Besides these field stream lines most densely concentrated in the cores, there exists other ones that fully crosses the stack of  layers, in the opposite direction relative to the sinking ones, never to come back again, like in an infinitely extended
solenoid. The skyrmion cores  are small, concentrated pockets of
opposite field with 9.4$\%$ and 5.3$\%$ of the total
cell area, for d and s wave respectively. In the pseudogap state
the Fermi surface breaks apart leading to the emergence of electron
pockets for hole-doped cuprates \cite{norman10}. This transformation
of the Fermi surface has been suggested to be a consequence of a new
periodicity that sets in the system \cite{taillefer09}. We find strong similarities between the electron-hole duality and the flow-counter flow circulating supercurrents of the skyrmion state.

The skyrmion's topological charge is obtained by integration over a
single layer, at $x_3=0$ for $\hat h = \vec h/\vert \vec h \vert$.
\begin{eqnarray}\label{skyrmion}
Q= \frac{1}{4\pi}\int_{L^2,\, x_3=0^+} \big (\frac{\partial \hat
h}{\partial x_1} \times \frac{\partial \hat h}{\partial x_2}
\big)\cdot \hat h \; d^2x,
\end{eqnarray}
and one obtains that Q=-2 and -1 for the d and s wave states,
respectively, in agreement with the plotted solutions of
fig.(\ref{currents}).  Clearly the time
reversal symmetry, ($\vec h \rightarrow -\vec h$), is broken by the
skyrmions.

\section{The number of carriers in the pseudogap}
The normal state of the underdoped
cuprate superconductors is unusual because it reveals a gap, the so-called
pseudogap~\cite{alloul89} that breaks the time-reversal
symmetry~\cite{kaminski02,xia08,he11}.
The interpretation of the pseudogap as a skyrmion state leads to an estimate of the number of carriers that condense in the pseudogap of the cuprate superconductors, $n_{pg}$.
We obtain here that $n_{pg} \sim 10^{-4} \,\mbox{nm}^{-3}$, or $0.01\%$ of the Cooper pair density in the cuprates.

Two-component OP theories are being considered to describe properties of  the high-temperature layered superconductors~\cite{hunte08, kohsaka08, agterberg98, moshchalkov09} and only a multi-component OP theory can yield a time
reversal broken state~\cite{sigrist91,takashi12}. According to Volovik and Gorkov a state described by an OP with
broken time reversal symmetry must have an accompanying magnetic
order ~\cite{volovik85}. For this reason there has
been an intense search for the accompanying magnetic order
associated to the pseudogap, and proposals have been made to explain
it, such as by C. M. Varma~\cite{varma06,fauque06,li11,bourges11},
based on microscopic orbital currents. Indeed polarized neutron
diffraction experiments~\cite{fauque06,li11} indicate a magnetic
order below the pseudogap, but NMR/NQR \cite{strassle08,strassle11}
and $\mu$SR \cite{macdougall08,sonier09} experiments set a very
restrictive limit to {\it the maximum magnetic field} inside the
superconductor, hereafter called $h_{max}$.

We obtain here that the gap density of the skyrmion state is $F_k \sim 0.1 \, h_{max}\,\mbox{meV.nm}^{-3}$,
where $h_{max}$ is given in Gauss. For the single-layer cuprates the inter-layer distance is $d \approx 1.0 nm$. To fix ideas we take that $h_{max}\sim
0.01\,\mbox{G}$, thus below the above experimental thresholds, and obtain for the gap density that $F_k \sim 10^{-3}
\,\mbox{meV.nm}^{-3}$. Recall that the BCS superconducting state which lies below the normal state and has a gap density of $F_{gap}=2\Delta n_{g}$, where $2\Delta$ is the energy required to break a single Cooper pair and $n_{g}$ represents the density of available Cooper pairs, namely $n_g=0.187\Delta n/E_F$, $n$ and $E_F$ being the electronic density and the Fermi energy, respectively.
Then one obtains that the gap density for metals is $F_{gap}\sim\,10^{-4} \,\mbox{meV.nm}^{-3}$
since $n \sim 0.1\,\mbox{nm}^{-3}$, $\Delta/E_F \sim 10^{-4}$ and
$2\Delta \sim 1.0 \,\mbox{meV}$. A similar estimate for the cuprates
gives that the gap density is $F_{gap}\sim 10 \,\mbox{meV.nm}^{-3}$,
considering that the gap is ten times larger than that of metals,
$2\Delta \sim 10 \,\mbox{meV}$, and $n_g\sim 1.0 \,\mbox{nm}^{-3}$,
since there are a few Cooper pairs~\cite{batlogg91} occupying the
coherence length volume, $\xi_{ab}^2\xi_c$, where $\xi_{ab}\sim 1.5
\,\mbox{nm}$, $\xi_c \sim 0.3 \, \mbox{nm}$. Our estimate of
$n_{pg}$ follows from the assumption that the skyrmion gap
(pseudogap) is equal to the superconducting gap. Then the large wavelength
limit of the kinetic energy (eq.(\ref{gap})) reveals the pseudogap
density for the cuprates, since $F_{k}\rightarrow 2\Delta_{pg} n_{pg}\sim 10^{-3}
\,\mbox{meV.nm}^{-3}$ and by assumption, $2\Delta_{pg} \sim 10 \,\mbox{meV}$.

\section{The theory}
We consider the scenario of a layered superconductor above its critical temperature T$_c$ such that its  free energy density is
widely dominated by the kinetic and field density energies over the
condensate energy, to the point that this last one can be neglected. Notice that the presence of a condensate energy will not affect the stability of the skyrmions, which is of topological nature.
Then the free energy density, is just the  sum of the
kinetic energy and field energy densities~\cite{doria10,alfredo13,doria12},
$F=F_k+F_f$,
\begin{eqnarray}
F_k = \langle \frac{\vert \vec{D}\Psi \vert^2}{2m} \rangle, \; \Psi
= \left(
\begin{array}{c} \psi_u \\ \psi_d \end{array} \right), \; \mbox{and} \; F_f = \langle\frac{\vec h^2}{8\pi}
\rangle,\label{fkf}
\end{eqnarray}
where $\langle \cdots \rangle \equiv \int \left(d^3x/V \right) \left
(\cdots \right)$ and $V$ is the bulk volume. There is minimal
coupling, $\vec D = (\hbar/i)\vec \nabla - (q/c)\vec A$, and the
local magnetic field is $\vec h = \vec \nabla \times \vec A$. We
solve the resulting variational equations for a set of equally
spaced superconducting layers, separated by $d$, with metallic
filling between them. The layers carry a superficial current $\vec
J_s$, which must also be determined. We do this by the first order
equation (FOE) method, which solves Amp\`ere's law exactly, $\vec
\nabla \times \vec h = 4\pi \vec J/c$, $\vec J =
(q/2m)\big(\Psi^{\dag}\vec D \Psi + c.c. \big)$, and the
Ginzburg-Landau equation, $\vec D^2\Psi=0$ approximately, meaning
that, instead, we solve exactly the so-called integrated equation,
$\langle \Psi^{\dag}\vec D^2\Psi\rangle=0$,  which can also be
expressed as $\langle \vert \vec D\Psi\vert^2-(\hbar/2)\vec
\nabla^2\vert\Psi\vert^2 \rangle=0$. The FOE method relies on the
following identity, which allows for a twofold view of the kinetic
energy~\cite{alfredo13}:
\begin{eqnarray}\label{kin2}
&&\frac{1}{2m}\left \vert\vec{D}\Psi\right\vert^2=
\frac{1}{2m}\left\vert\vec{\sigma}\cdot\vec{D}\Psi\right\vert^2+\frac{\hbar
q}{2mc}\vec{h} \cdot\Psi^\dag\vec{\sigma}\Psi- \nonumber
\\&&-\frac{\hbar}{4m}\vec \nabla\left[\Psi^\dag\left(\vec \sigma \times
\vec D\right)\Psi+c.c. \right].
\end{eqnarray}
From it an equivalent, but distinct, formulation of the current
density is obtained,
\begin{equation}
\vec J=\frac{q}{2 m}\left[\Psi^\dag\vec \sigma\left
(\vec{\sigma}\cdot\vec{D}\Psi\right )+ c.c.\right ]- \frac{\hbar
q}{2 m}\vec \nabla \times \left(\Psi^\dag \vec{\sigma}\Psi\right).
\label{curr2}
\end{equation}
Imposing that the order parameter satisfies
\begin{eqnarray}
\vec{\sigma}\cdot\vec{D}\, \Psi=0, \label{foe1}
\end{eqnarray}
leads to the exact determination of the local field from Amp\`ere's
law,
\begin{eqnarray}
\vec h = \vec C - 4\pi\mu_B\Psi^\dag  \vec{\sigma}\Psi, \label{foe2}
\end{eqnarray}
where $\mu_B=\hbar q/2mc$ is Bohr's magneton. Thus for fields
that satisfy the FOE, eq.(\ref{kin2}) becomes:
\begin{eqnarray}\label{kin3}
&&\frac{1}{2m}\left \vert\vec{D}\Psi\right\vert^2=\mu_B\left(\vec C
- 4\pi\mu_B\Psi^\dag  \vec{\sigma}\Psi \right)\cdot\Psi^\dag
\vec{\sigma}\Psi+ \nonumber
\\
&&+\frac{\hbar^2}{4m}\nabla^2\left(\Psi^\dag\Psi\right)
\end{eqnarray}
Introducing eq.(\ref{kin3}) into the integrated equation gives that
$\langle (\vec C - 4\pi \mu_B \Psi^{\dag}\vec \sigma \Psi)\cdot
\Psi^{\dag}\vec \sigma \Psi \rangle =0$, which is exactly solved for
the following choice of integration constant:
\begin{eqnarray}
\vec C = 4\pi\mu_B \langle \Psi^\dag \vec \sigma \Psi \rangle
\frac{\langle \left(\Psi^\dag \vec \sigma \Psi
\right)^2\rangle}{\langle \Psi^\dag \vec \sigma \Psi \rangle^2}.
\label{cconst}
\end{eqnarray}
Thus one obtains that,
\begin{eqnarray}
&& F_k=\frac{\hbar^2}{4m}\langle \vec \nabla^2\left( \Psi^{\dag}\Psi\right)\rangle, \; \mbox{and} \label{fkg}\\
&& F_f = 2\pi\mu_B^2\langle \left(\Psi^\dag \vec \sigma \Psi
\right)^2\rangle\left[ \frac{\langle \left(\Psi^\dag \vec \sigma
\Psi \right)^2\rangle}{\langle \Psi^\dag \vec \sigma \Psi
\rangle^2}-1. \right] \label{fenop}
\end{eqnarray}
Eqs.(\ref{foe1}) and (\ref{foe2}) are nonlinear and must be
solved iteratively. The first one determines $\Psi$ while the second
one, $\vec h$. We solve them in the lowest order approximation, that
is, firstly $\Psi$ is obtained from eq.(\ref{foe1}) in the absence
of $\vec h$, and, next, $\vec h$ is determined from eq.(\ref{foe2}),
using the known OP solution. The success of this lowest
approximation relies on the fact that $\vec h$ must be very weak,
such that no further iterations of the FOE are needed. For the case
of a single layer, the solution of
$\vec{\sigma}\cdot\vec{\nabla}\Psi=0$ is, for $x_3 \neq 0$,
\begin{eqnarray}\label{sls}
\Psi = \sum_{\vec k \neq 0} c_{\vec k} \; e^{-k\vert x_3 \vert}
e^{i\vec k \cdot \vec x} \left(
\begin{array}{c} 1 \\ -i\frac{k_{+}}{k}\frac{x_3}{\vert x_3 \vert} \end{array}\right)
\end{eqnarray}
where $k_\pm = k_1 \pm i k_2$. The space-time symmetries are broken,
namely, the reflection symmetry, $x_3$ $\rightarrow$ $-x_3$, and
also the time reversal symmetry, $k_{+}\rightarrow k_{-}$. Although
these features reflect our particular choice of a basis, they are
intrinsic to the solution and cannot be removed from it. Notice that
the density $\Psi^{\dag}\Psi$ and $\Psi^{\dag}\sigma_3\Psi$ are
continuous across the layer, and so is $h_3$, according to
eq.(\ref{foe2}). Nevertheless $\Psi^{\dag} \sigma_{1}\Psi$ and
$\Psi^{\dag}\sigma_{2}\Psi$ are discontinuous across the layer, and
so is $\vec h_{\parallel} \equiv h_1\hat x_1 + h_2\hat x_2$, and,
consequently, there is a superficial current density,
\begin{eqnarray}\label{curr}
\vec J_s = - 2c\mu_B \, \hat x_3 \times \Psi^\dag (0^+)
\vec{\sigma}\Psi(0^+).
\end{eqnarray}
We next solve $\vec{\sigma}\cdot\vec{\nabla}\Psi=0$ for a stack of
layers under the simplifying assumption that all layers are
identical:
\begin{eqnarray}\label{mls}
\Psi = \sum_{\vec k \neq 0} c_{\vec k} \; \frac{e^{i\vec k \cdot
\vec x}}{\sinh\left(k d/2\right)} \left(
\begin{array}{c} \cosh \left[k\left(x_3-d/2\right) \right] \\ i\frac{k_{+}}{k} \sinh \left[k\left(x_3-d/2\right)
\right]\end{array}\right).\nonumber \\ \label{op3d}
\end{eqnarray}
The field $\vec h$ is obtained analytically by introducing
eq.(\ref{mls}) into eq.(\ref{cconst}) firstly, and next to
eq.(\ref{foe2}). There is a fundamental difference between {\it
single} and {\it multiple} layer solutions described by
eq.(\ref{sls}) and (\ref{mls}), respectively. For a single layer
$\langle \vec h \rangle = 0$, since $\langle \Psi^{\dag} \vec \sigma
\Psi \rangle=0$. In fact for the single layer eq.(\ref{cconst}) does
not apply to determine $\vec C$. For a stack of layers  $\langle
\Psi^{\dag} \sigma_3 \Psi \rangle =\sum_{\vec k \neq 0} \vert
c_{\vec k}\vert^2/\sinh^2\left( kd/2\right)$,  and  $\langle
\Psi^{\dag} \vec \sigma_{\parallel} \Psi \rangle=0$. Thus the stack
of layers behaves similarly to a solenoid, which renders $\langle
h_3 \rangle \ge 0$, and only $\langle \vec h_{\parallel}\rangle=0$.
The low component of eq.(\ref{mls}) vanishes in the middle plane
($x_3=d/2$), and there $\Psi^{\dag} \vec \sigma_{_{\parallel}}
\Psi|_{x_3=d/2} =0$. Physically this means that $\vec h_{\parallel}$
has opposite signs in $0<x_3<d/2$ and $d/2<x_3<d$, and consequently
there is a non-zero $\vec J_s$ in the layers. $F_k$ and $F_f$ are
well defined only outside the layer, thus excluding $x_3 = 0$. The
same holds for the multilayer solution of eq.(\ref{mls}), limited to
$0<x_3 < d$.

\begin{figure}[htb]
\includegraphics[width=\linewidth]{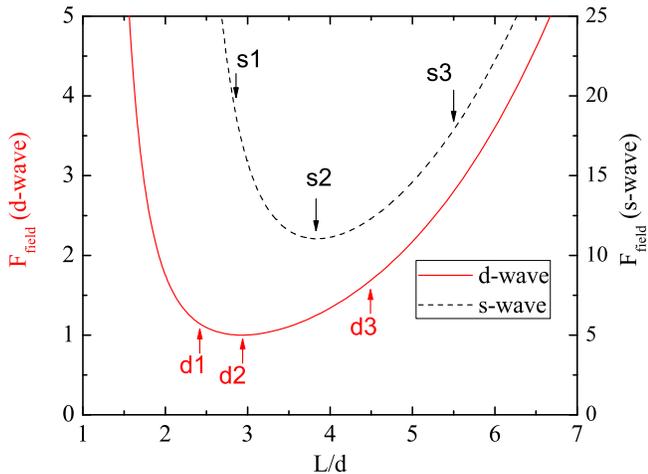}
\caption{\label{magener}(color online) The magnetic energy between
layers versus $L/d$ is shown here. Three  $L/d$ ratios are shown for
d-wave, 2.42 (d1), 2.94 (d2) and 4.49 (d3), and also for s-wave,
2.86 (s1), 3.83 (s2), and 5.50 (s3). The d2 and s2 points are the
minimum of d and s wave curves, respectively. The curves are
normalized to the minimum of the d-wave magnetic energy.}
\end{figure}

\begin{figure}[htb]
\includegraphics[width=\linewidth]{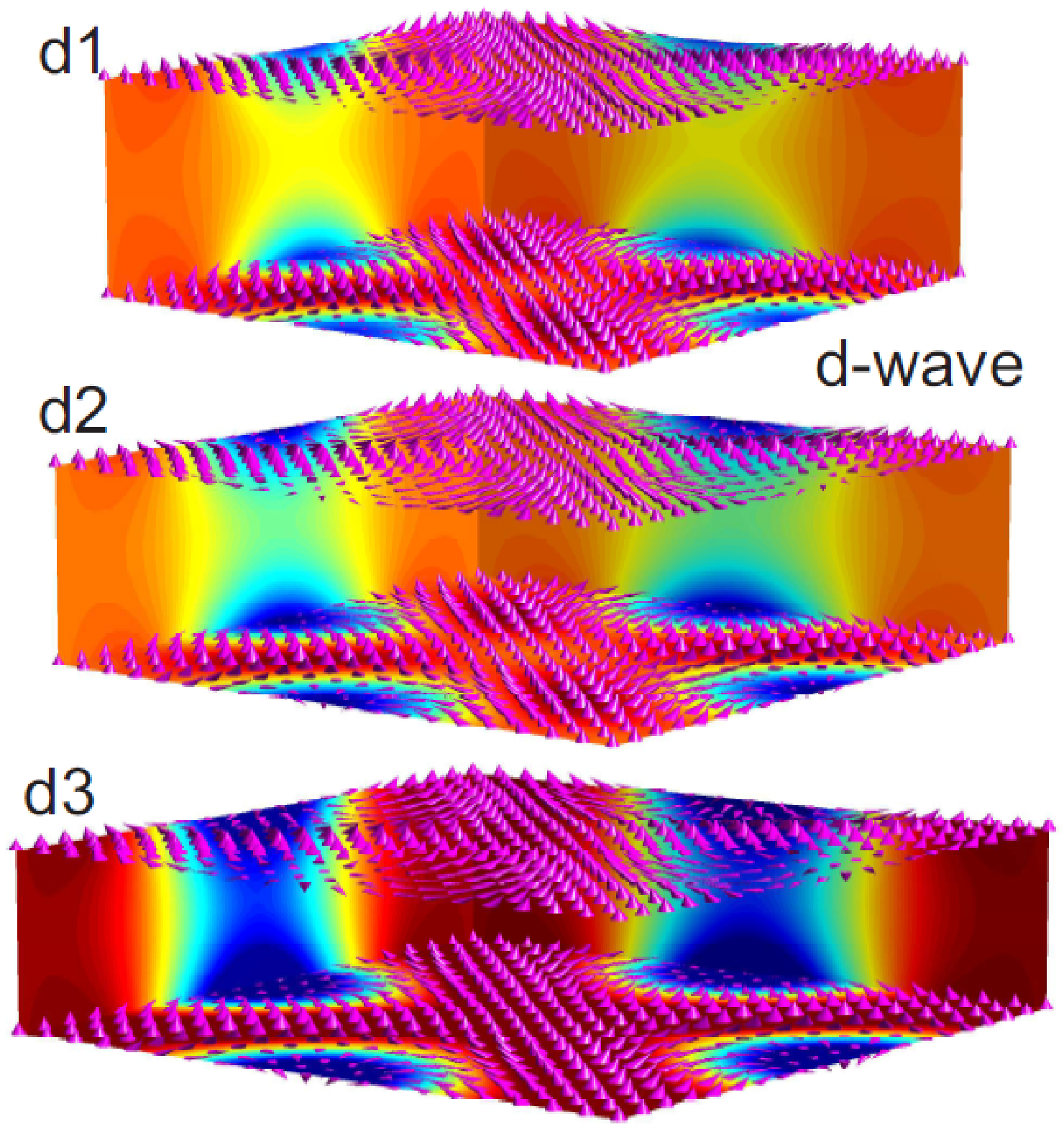}
\caption{\label{h3d}(color online) The d-wave local magnetic field
component, perpendicular to the layers, $h_3$, is shown in colors
(blue negative, green zero and red positive) at the walls of the
$dL^2$ unit cell. The (cyan) cones depict the local magnetic field
$\vec h$ infinitesimally {\it below} (top cones) and {\it above}
(bottom cones) a layer.}
\end{figure}

\begin{figure}[htb]
\includegraphics[width=\linewidth]{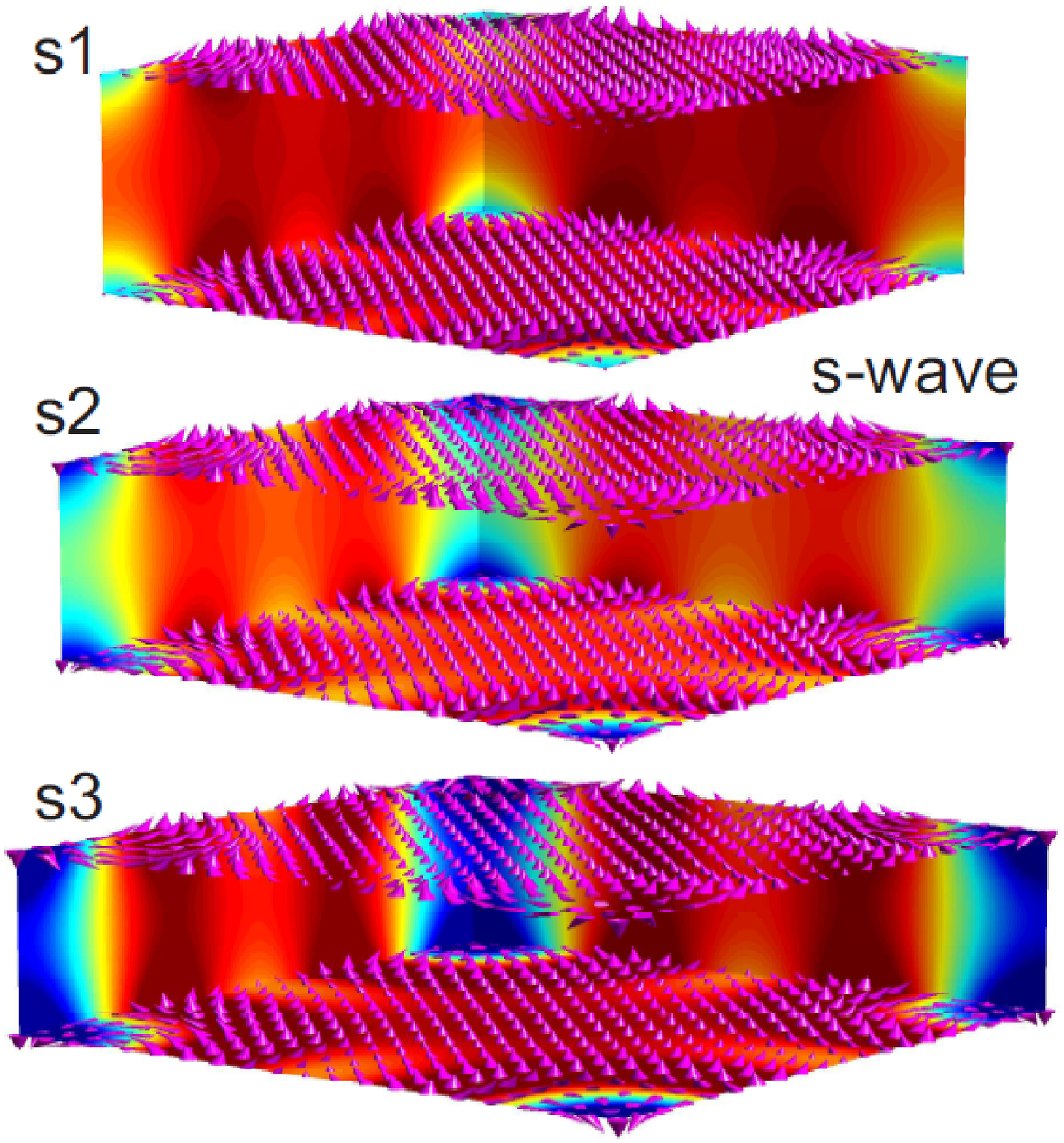}
\caption{\label{h3s}(color online) The s-wave local magnetic field
component perpendicular to the layers, $h_3$, is shown in colors
(blue negative, green zero and red positive) at the walls of the
$dL^2$ unit cell. The (cyan) cones depict the local magnetic field
$\vec h$ infinitesimally {\it below} (top cones) and {\it above}
(bottom cones) the layer.}
\end{figure}

The gap density of the skyrmion state follows from eqs.(\ref{fkg})
and (\ref{op3d}), which give that,
\begin{eqnarray}\label{gap}
F_k=\frac{(h/d)^2}{\pi^2m}\sum_{\vec k \neq 0} \vert c_{\vec
k}\vert^2 \frac{kd/2}{\tanh\left(kd/2\right)}.
\end{eqnarray}
The gap is not very sensitive to the small $k$ regime as shown by the two limits of the function $f(z)=z/\tanh(z)$, $z\equiv kd/2$: $f(z \rightarrow 0) \rightarrow 1$, and $f(z=1) \approx 1.3$. Nevertheless the region near to $z=1$ matters, because, as shown below, there lives the skyrmion state, which is stable.

\section{The d- and s-wave models}
Here we select the coefficients $c_{\vec k}$ in eq.(\ref{op3d}) that
represent  $s$ and $d$ waves, which are well defined values of the
angular momentum perpendicular to the layers, namely, of the
operator $L_3 = (\hbar/i)\left ( k_1\partial/\partial k_2
-k_2\partial/\partial k_1 \right )$. The $s$ and $d$ states
correspond to the eigenvalues $m=0$ and $m=\pm 2$, respectively of
$L_3 c_{\vec k} = \pm m \hbar c_{\vec k}$, $c_{\vec k}
=\left(k_{\pm}\right)^m$. Therefore the $s$ and $d$ states
correspond to $c^s_{\vec k}=\psi_0$ and $c^d_{\vec k}
=\psi_0\left(k_{+}^2 + k_{-}^2 \right)/2k^2$, where $\psi_0^2$ is
the density that determines $h_{max}$ according to eq.(\ref{foe2}).
We choose $c_{\vec k}$ for the $d$ wave state to be an equal
admixture of $m=2$ and $m=-2$ states. Then they are eigenstates of
$L_3^2$, and not of $L_3$. Although the coefficients $c^{d,s}_{\vec
k}$ are common to both components of $\Psi$, $\psi_d$, is in a
different $L_3^2$ state than $\psi_u$, because $k_{+}$ adds an extra
angular momentum besides breaking  the time reversal symmetry.
For simplicity only the lowest fourier terms are included, and so,
$\vec k = 2\pi \vec n/L$, $\vec n = n_1 \hat x_1 + n_2 \hat x_2$, where,
$n_i=-1,0,1$, $i=1,2$. For the totally inhomogeneous state
$n_1=n_2=0$ is excluded.

The  very weak $\vec h$ of the skyrmion state yields a negligible $F_f$ as compared to $F_k$, and consequently, to the gap  density of eq.(\ref{gap}).
The search of the optimal unit cell ratio, $L/d$, that
minimizes $F_f$ also determines the region of existence of the skyrmion state, which lives near this minimum.
According to eq.(\ref{fenop}) $F_f$ plunges to a
minimum within a small $L$ range window, around $L/d \sim 3$ and $4$
for $d$ and $s$ wave symmetries, respectively, as shown in
fig.(\ref{magener}). The values d2 and s2 correspond to such minima
while the other points (d1, d3) and (s1, s3) are arbitrary
selections chosen below and above these minima. Fig.(\ref{magener})
shows that the d-wave state (red curve, left scale) has almost ten
times less stored magnetic energy than the s-wave state (black
curve, right scale). Figs. (\ref{h3d}) and (\ref{h3s}) show that the
region around the center of the skyrmions are pockets of $h_3<0$,
thus in opposite direction to the majority of the unit cell, which
features $h_3>0$. The configurations of figs. (\ref{h3d}) and
(\ref{h3s}) correspond to the three selected $L/d$ ratios  of
fig.{\ref{magener}}. The same red to blue scale applies to each of
the three displayed cases. The (cyan) cones of figs.(\ref{h3d}) and
(\ref{h3s}) depict $\vec h$ slightly below and above a layer,
respectively. Because all layers are equivalent, the bottom plane of
(cyan) cones shows $\vec h$ slightly above a layer. Conversely, the
top plane of (cyan) cones shows $\vec h$ slightly below a layer. In
this way the cones around the center of the skyrmions clearly show
the discontinuity of $\vec h_{\parallel}$, and so, the presence of
$\vec J_s$ in the blue regions. Regions with a dominant $h_3$
component, have very weak $\vec J_s$, because $h_3$ is continuous
across the layer. These regions take most of the unit cell,
specially its center, where (cyan) cones point upward, but there is
also the center of the skyrmions, where $h_3<0$. Notice that the
configurations (d2, s2) contain more green color ($h_3=0$) than the
(d1, s1), and (d3, s3) ones, respectively, in agreement to the fact
that there the lowest magnetic energy is reached. Although (d3, s3)
are the configurations with most intense fields (red, $h_3>0$; blue,
$h_3<0$), their magnetic energies are larger than that of (d2, s2).
The same holds for (d1, s1), which are the configurations with less
$h_3<0$ regions as compared to the others.

The present FOE follow from a more general Lichnerowicz-Weitzenb\"ock
formula~\cite{alfredo13}, that treats superconductivity in presence of spatial inhomogeneities caused by external spin and charge degrees of
freedom that bring curvature and torsion to space, as in the geometrical approach of \'Elie Cartan~\cite{katanaev92}.
We hope that the present macroscopic approach will bring some understanding to the microscopic mechanism of pairing in the underdoped regime of the cuprates~\cite{marchetti11}.

\section{Conclusion}
Using the first order equations we show that the two-component order parameter layered superconductor has a
topologically stable inhomogeneous state with a gap above the homogeneous ground state. It
breaks the time reversal symmetry and produces a very  weak inner
magnetic field. It is made of skyrmions, whose cores form pockets of
negatively oriented magnetic field in the layers. For all the above reasons we suggest that the pseudogap is a skyrmion state.

\acknowledgments
Alfredo Vargas Paredes and Mauro M. Doria acknowledge CNPq for financial support.


\begin{thebibliography}{0}

\bibitem{schmalian10} J. Schmalian \textit{Modern Physics Letters B} \textbf{24} 2010 2679.

\bibitem{abrikosov57} A. A. Abrikosov \textit{Soviet Physics JETP} \textbf{5} 1957 1174.

\bibitem{bogomolny76} E. B. Bogomolny \textit{Sov. J. Nucl. Phys.} \textbf{24} 1976 449.

\bibitem{seiberg94} N. Seiberg and E. \textit{Witten Nuclear Physics B} \textbf{426} 1994 19.


 \bibitem{doria10} M. M. Doria, A. R. de C. Romaguera, and F. M. Peeters \textit{Europhysics Letters} \textbf{92} 2010 17004.

\bibitem{alfredo13} A. A. Vargas-Paredes, M. M. Doria, and J. A. H. Neto \textbf{Journal of Mathematical Physics} \textbf{54} 2013 013101.

 \bibitem{babaev02} Babaev, Egor and Faddeev, Ludvig D. and Niemi, Antti J. \textit{Phys. Rev. B} \textbf{65} 2002 100512.

\bibitem{garaud12}  Garaud, Julien and Babaev, Egor  \textbf{Phys. Rev. B} \textbf{86} 2012 060514.

\bibitem{skyrme62}  T. Skyrme  \textit{Nuclear Physics} \textbf{31} 1962  556.

\bibitem{kalbermann91} G. K\"albermann, G. Pari, and J. M. Eisenberg  \textit{Phys. Rev. C} \textbf{44} 1991 899.

\bibitem{schmeller95}  A. Schmeller, J. P. Eisenstein, L. N. Pfeiffer, and K. W. West  \textit{Phys. Rev. Lett.} \textbf{75} 1995 4290.
	
\bibitem{volovik12}  G. E. Volovik and M. Krusius  \textit{Physics} \textbf{5} 2012 130.
	
\bibitem{walmsley12}  P. M. Walmsley and A. I. Golov  \textit{Phys. Rev. Lett.} \textbf{109} 2012 215301.
	
\bibitem{marcelo11}  I. Rai\u{c}evi\'c et al.  \textit{Phys. Rev. Lett.} \textbf{106} 2011 227206.
	
\bibitem{yu10}  X. Z. Yu at al. \textit{Nature} \textbf{465} 2010 901.
	
\bibitem{pfleiderer04}  C. Pfleiderer et al. \textit{Nature} \textbf{427} 2004 227.
	
\bibitem{munzer10}  W. M\"unzer at al.  \textit{Phys. Rev. B} \textbf{81} 2010 041203.

\bibitem{edinardo13}  Alfredo A. Vargas-Paredes, Marco Cariglia, Mauro M.Doria, Edinardo I.B. Rodrigues, and A.R.C. Romaguera  \textit{J Supercond Nov Magn} 2013 1.
  
\bibitem{roeser08}  H. Roeser, F. Hetfleisch, F. Huber, M. von Schoenermark, M. Stepper, A. Moritz, and A. Nikoghosyan  \textit{Acta Astronautica} \textbf{62} 2008 733.


\bibitem{tranquada08}  J. M. Tranquada, G. D. Gu, M. H¨ucker, Q. Jie, H.-J. Kang, R. Klingeler, Q. Li, N. Tristan, J. S. Wen, G. Y. Xu, Z. J. Xu, J. Zhou, and M. v. Zimmermann  \textit{Phys. Rev. B} \textbf{78} 2008 174529.

\bibitem{norman10}  M. R. Norman  \textit{Physics} \textbf{3} 2010 86.
	
\bibitem{taillefer09}  L. Taillefer  \textit{Condensed Matter} \textbf{21} 2009 164212.

\bibitem{hunte08}  F. Hunte et al.  \textit{Nature} \textbf{453} 2008 903.
	
\bibitem{kohsaka08}  Y. Kohsaka et al.  \textit{Nature} \textbf{454} 2008 1072.
	
\bibitem{agterberg98}  D. F. Agterberg  \textit{Phys. Rev. Lett.} \textbf{80} 1998 5184.

\bibitem{moshchalkov09}  V. Moshchalkov et al.  \textit{Phys. Rev. Lett.} \textbf{102} 2009 117001.

\bibitem{sigrist91}  M. Sigrist and K. Ueda  \textit{Rev. Mod. Phys.} \textbf{63} 1991 239.
	
\bibitem{takashi12}  T. Yanagisawa, Y. Tanaka, I. Hase, and K. Yamaji  \textit{Journal of the Physical Society of Japan} \textbf{81} 2012 024712.
	
\bibitem{volovik85}  G. E. Volovik and L. P. Gor'kov  \textit{Sov. Phys. JETP} \textbf{61} 1985 843.
	
\bibitem{alloul89}  H. Alloul, T. Ohno, and P. Mendels  \textit{Phys. Rev. Lett.} \textbf{63} 1989 1700.

\bibitem{xia08}  J. Xia et al.  \textit{Phys. Rev. Lett.} \textbf{100} 2008 127002.
	
\bibitem{he11}  R. H. He at al.  \textit{Science} \textbf{331} 2011 1579.
	
\bibitem{kaminski02}  A. Kaminski et al.  \textit{Nature} \textbf{416} 2002 610.

\bibitem{varma06}  C. M. Varma  \textit{Phys. Rev. B} \textbf{73} 2006 155113.
	
\bibitem{fauque06}  B. Fauqu\'e et al.  \textit{Phys. Rev. Lett.} \textbf{96} 2006 197001.
	
\bibitem{li11}  Y. Li et al.  \textit{Phys. Rev. B} \textbf{84} 2011 224508.

\bibitem{bourges11}  P. Bourges and Y. Sidis  \textit{Comptes Rendus Physique} \textbf{12} 2011 461.
	
\bibitem{strassle08}  S. Str\"assle, J. Roos, M. Mali, H. Keller, and T. Ohno  \textit{Phys. Rev. Lett.} \textbf{101} 2008 237001.
	
\bibitem{strassle11}  S. Str\"assle, B. Graneli, M. Mali, J. Roos, and H. Keller \textit{Phys. Rev. Lett.} \textbf{106} 2011 097003.
	
\bibitem{macdougall08}  G. J. MacDougall et al.  \textit{Phys. Rev. Lett.} \textbf{101} 2008 017001.
	
\bibitem{sonier09}  J. E. Sonier et al.  \textit{Phys. Rev. Lett.} \textbf{103} 2009 167002.
	
\bibitem{batlogg91} B. Batlogg  \textit{Physics Today} \textbf{44} 1991 44.
	
\bibitem{doria12}  M. M. Doria, A. A. Vargas-Paredes, and J. A. Helayel-Neto  \textit{Modern Physics Letters B} \textbf{26} 2012 1230005.

\bibitem{katanaev92}  M.O. Katanaev and I.V. Volovich  \textit{Ann. Phys.} \textbf{216} 1992 1.

\bibitem{marchetti11}  P. A. Marchetti, F. Ye, Z. B. Su and L. Yu  \textit{Europhysics Letters} \textbf{93} 2011 57008.
	
\end{thebibliography}
\end{document}